# Landau Damping with Electron Lenses in Space-Charge Dominated Beams

Y. Alexahin, A. Burov, V. Shiltsev

*Fermi National Accelerator Laboratory, PO Box 500, Batavia, IL 60510, USA*

*Abstract*

Progress on the Intensity Frontier of high energy physics critically depends on record high intensity charged particles accelerators. Beams in such machines become operationally limited by coherent beam instabilities, particularly enhanced in the regime of strong space charge (SC). Usual methods to control the instabilities, such as octupole magnets, beam feedback dampers and employment of chromatic effects, become less effective and insufficient. In [1] it was proposed to employ electron lenses for introduction of sufficient spread in particle oscillation frequencies needed for beam stabilization and in [2] it was shown that electron lenses are uniquely effective for Landau damping of transverse beam instabilities in high energy particle accelerators and their employment does not compromise incoherent (single particle) stability, dynamic aperture and the beam lifetime. Here we consider an important issue of effectiveness of the Landau damping with electron lenses in space-charge dominated beams and demonstrate that the desired stability can be assured with proper choice of the electron beam parameters and current distributions.

1. INTRODUCTION

There are many impedance-driven collective instability phenomena in high intensity charged particle beams [3, 4]. These instabilities commonly limit the single-bunch and total beam intensities and constitute a major performance-limiting phenomenon in, e.g., proton synchrotrons. Suppression of the collective instabilities can be obtained via the stabilizing effect of Landau damping [5] by spreading the spectrum of incoherent frequencies $\omega_{x,y}$ to overlap it with the frequencies of the unstable collective modes, thus allowing absorption of the collective energy by the resonant particles. Thus far, commonly used are octupole magnets with the transverse magnetic fields on beam's axis of $B_x + iB_y = O_3(x + iy)^3$ which generate the betatron oscillation frequency shifts proportional to the squares of particle's amplitudes [6]. Damping by octupoles has several drawbacks: first of all, the corresponding frequency spread $\delta\omega$ scales with beam energy increase as $1/E^2$ due to increasing rigidity and smaller beam size (the rms amplitude of the particle's betatron oscillations $(x,y)=(2J_{x,y}\beta_{x,y})^{1/2}\cos(\omega_{x,y}t+\phi_{x,y})$, where $\beta_{x,y}$ and $\phi_{x,y}$ are corresponding location-dependent focusing optics beat-functions and phases), hence, one needs to increase strength of these magnets accordingly. Secondly, strong octupoles significantly reduce machine's dynamic aperture. Another method involves beam-based feedback system which suppresses coherent motion of the beam or bunch centroid. Though generally effective, such feedback systems which act only on the modes with non-zero dipole moment, leaving the multitude of other "head-tail" modes unsuppressed [7]. Electron lenses [8] were shown to provide effective Landau damping [1, 2] mechanism free of all the above listed drawbacks of other methods.

The Lorenz force on ultra-relativistic proton due to low energy electron beam with current density distribution $j_e(r)$ is equal to [8]:

$$e\left(E_r + \beta B_\theta\right) = \frac{4\pi e(1+\beta_e)}{\beta_e c} \frac{1}{r} \int_0^r j_e(r')r'dr' . \tag{1}$$

It is diminishing at large radii $r$ as $\sim 1/r$ and, therefore, the corresponding betatron frequency shift decreases as an inverse square of the proton's amplitude of the betatron oscillations. In the case of the round Gaussian-profile electron beam with the rms transverse size $\sigma_e$, the amplitude dependent tune shift $\delta\omega_x / \omega_0 \equiv \delta\nu_x$ is equal to [9]:

$$\delta\nu_x(\kappa_x,\kappa_y) = 2\delta\nu_{max} \int_0^{1/2} \frac{I_0(\kappa_x u) - I_1(\kappa_x u)}{\exp(\kappa_x u + \kappa_y u)} I_0(\kappa_y u)du;$$

$$\kappa_{x,y} = \frac{a_{x,y}^2}{2\sigma_e^2}; \quad \delta\nu_{max} = \frac{I_e}{I_A} \frac{m_e}{m_p} \frac{\sigma_x^2}{\sigma_e^2} \frac{L_e}{4\pi\varepsilon_n} \frac{1+\beta_e}{\beta_e}, \tag{2}$$

and it vanishes at large $a_{x,y}/\sigma_e$, as it is shown in Fig.1. Here the tune shift refers to the shift of the betatron oscillation frequency in the units of the revolution frequency in the ring $\omega_0$, $I_{0,1}(x)$ are the modified Bessel functions, $\beta_e$ is the electron relativistic factor, $L_e$ is the length of the electron beam, $I_e$ is the electron current, $I_A$ =17kA is the Alfven current, $\varepsilon_n$ is the normalized rms emittance, $\sigma_x$ is the beam rms size, $m_e$ and $m_p$ are electron and proton masses. As shown in [2], the electron lenses generate the spread of the betatron frequencies of about 0.14 $\delta\nu_{max}$ which is usually sufficient for the Landau damping in the high energy beams.

Below we consider application of the e-lenses for Landau damping when SC has to be taken into account.

## 2. LANDAU DAMPING, ITS LOSS AND RESTORATION

Since space charge shifts single particle tunes away from the coherent tune, it may suppress Landau damping, up to its complete loss [10, 15, 16]. As nonlinear optical elements, electron lenses can help to restore the damping [11].

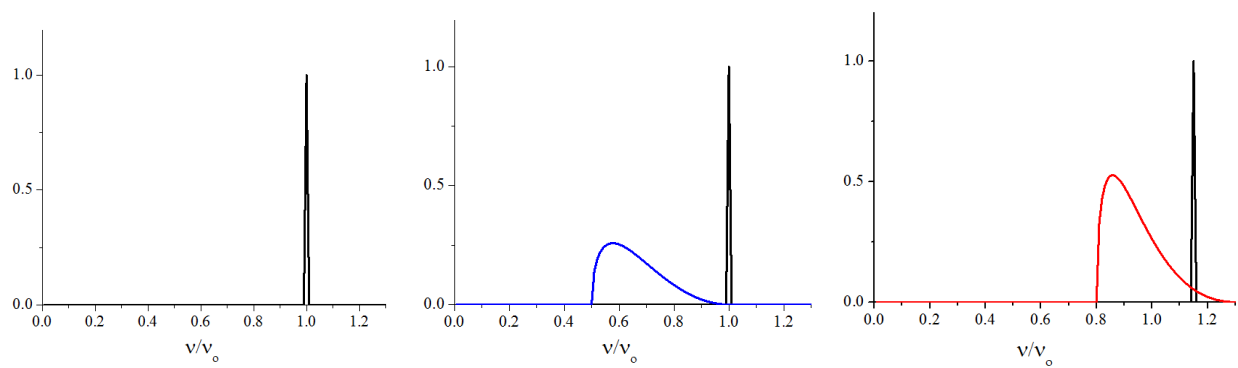

Figure 1: Illustrative dynamics of the spectra of coherent and incoherent betatron oscillations: a) left plot – in the absence of space charge forces; b) center – with strong space charge effect, but no electron lens, blue line – for incoherent frequencies, black one – for coherent; c) right - with an electron lens and in the presence of strong space charge effect, red line – for incoherent frequencies, black one – for coherent.

Schematically, the effect is illustrated in Fig. 1: in the absence of space charge the coherent and incoherent beam oscillations occur at the same frequency; space charge forces shift the spectrum of the incoherent oscillation frequencies to the left leaving no particles to absorb the power of the coherent waves, as needed for Landau damping; the electron lens with proper current density distribution differentially moves to the right (focusing effect) both the coherent tune and the incoherent oscillation spectrum such that the two overlap, resulting in the so-much sought Landau damping. The reason why the right edge of the incoherent spectrum is shifted more than the coherent line is that the coherent tune reacts on the beam-averaged value of the lens focusing, while the incoherent edge reflects the maximal tune shift. This qualitative picture, though, lacks quantitate analysis of the oscillation power spectrum and in-depth understanding of the efficiency of the power transfer from coherent to incoherent motion.

In bunched beams the reduction of the space charge tune shift for particles with large transverse amplitudes makes it possible to use a hollow electron lens. Such lens has a relaxed tolerance on the offsets, it affects mainly particles with intermediate betatron amplitudes and therefore only slightly if not at all reduces the dynamic aperture.

Figure 2 shows histograms of the spectral density of transverse oscillations in a bunched beam after receiving a kick in the presence of a hollow electron lens obtained by method described in the next Section. The lens transverse dimensions were chosen so that the maximum tuneshift, $\delta\nu_{max}$, was reached at oscillations amplitude of $3.4\sigma_\perp$, $\sigma_\perp$ being the beam transverse r.m.s. size. With increasing $\delta\nu_{max}$ the spectral peak is shifting (but not as much) and widens testifying of increasing Landau damping. The synchrotron tune for this example was $\nu_s = 0.02 \cdot \xi$ with $\xi = Q_{SC}$ being the maximum absolute value of the space charge tuneshift.

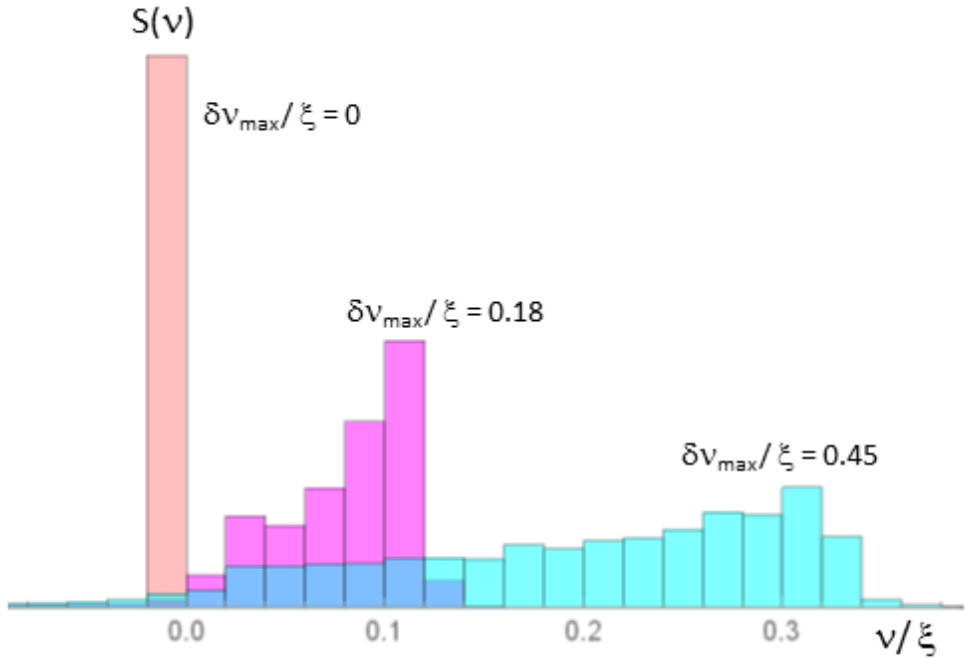

Figure 2. Spectral density of transverse oscillations in a bunch with space charge at indicated values of the maximum tuneshift due to a hollow electron lens.

3. EIGENMODES IN SPACE-CHARGE DOMINATED BEAM

To address the problem of Landau damping in SC-dominated bunched beams a rigorous approach based on the eigenmodes of Vlasov equation was developed which will be discussed in detail elsewhere. Here we limit ourselves to a simple case of horizontal oscillations in flat beams so that all dependencies on vertical amplitude can be ignored and consider Gaussian unperturbed beam distribution in all coordinates. Also, we neglect the effect of dispersion.

Using normalized action variable for horizontal motion $J_x = a_x^2/2\sigma_x^2$ and dimensionless longitudinal coordinate $\tau = z/\sigma_z$ and relative momentum deviation $\upsilon = (p - p_0)/\sigma_p$ we can write for the perturbed distribution function

$$F_1 = e^{i\psi_x - J_x/2 - (\tau^2 + \upsilon^2)/2} f(J_x, \tau, \upsilon; \theta)/(2\pi)^2 \varepsilon_x \varepsilon_z + c.c. \tag{3}$$

with $\psi_x$ being canonical angle variable and $\theta$ being the generalizing azimuth serving as the independent variable. Vlasov equation takes the form

$$i\frac{\partial}{\partial \theta} f = \hat{A} f \tag{4}$$

with the part of operator $\hat{A}$ describing external focusing and space charge being

$$\hat{A}_0 f = -i\nu_s (\upsilon \frac{\partial}{\partial \tau} - \tau \frac{\partial}{\partial \upsilon}) f + \nu_x^{(ext)} f + e^{-\tau^2/2} \left[ \nu_x^{(SC)}(J_x) f + \frac{1}{\sqrt{2\pi}} \hat{G} \int_{-\infty}^{\infty} e^{-\upsilon'^2/2} f(J_x, \tau, \upsilon'; ) d\upsilon' \right] \tag{5}$$

In the considered case of a flat beam

$$\nu_x^{(SC)} = -\xi \cdot (1 - e^{-J_x})/J_x, \quad \hat{G} f = \xi \cdot \int_0^{\infty} e^{-(J_x + J_x')/2} \sqrt{\frac{\min(J_x, J_x')}{\max(J_x, J_x')}} f(J_x') dJ_x' \tag{6}$$

The proposed method involves determination of spectrum of operator $\hat{A}_0$. Its eigenfunctions can be sought as expansion in some smooth basis functions $\Phi_l$ of longitudinal variables:

$$\underline{V}_m(J_x, \tau, \upsilon) = \sum_{k,l} V_{m;k,l}(J_x) \Phi_k(\upsilon) \Phi_l(\tau) \tag{7}$$

Generally it is convenient to choose functions $\Phi_l(\tau)$ being orthogonal with the longitudinal profile serving as weight function $w$; for a Gaussian beam $w = e^{-\tau^2/2}$. The choice

$$\Phi_k(\tau) = \sqrt{\frac{2k+1}{\sqrt{2\pi}}} P_k[\mathrm{erf}(\frac{\tau}{\sqrt{2}})], \quad k = 0, 1, \ldots, \infty \tag{8}$$

where $P_k(u)$ are the Legendre polynomials, proved to be quite advantageous. These functions satisfy the boundary condition $\Phi_l(\pm\infty)=0$ established in [15].

In action variable $J_x$ operator $\hat{A}_0$ has an integral part and a multiplicative part. In the result its spectrum may contain a discrete (point) set as well as continuum [17], the latter covering the range of incoherent tunes. The corresponding eigenmodes sometimes are referred to as Van Kampen modes. Discrete eigenmodes with tunes separated from the Van Kampen continuum can become unstable (acquire imaginary part of the tune) with addition of an infinitesimal impedance part to $\hat{A}$. Therefore a qualitative necessary condition of stability is the absence of discrete spectrum.

This condition can be visualized with the help of spectral coefficients describing the projection of eigenmodes on a pickup and their excitation by a dipole kick varying longitudinally as $\Phi_l(\tau)$

$$c_{m;l} = \int_0^\infty \mathcal{R}(J_x) v_{m;0,l}(J_x) dJ_x \qquad (9)$$

where $\mathcal{R}(J_x) = \sqrt{J_x} e^{-J_x/2}$ is function describing horizontal "rigid-slice" motion.

Figure 2 shows histograms of $l=0$ head-tail mode spectral coefficient, $c_{m;0}^2$, vs the eigenvalue normalized by $\xi$. The pink bar represents the discrete mode in absence of external nonlinearities, $\nu=0$, its transverse dependence is given by $\mathcal{R}(J_x)$ hence $c_{0;0}=1$. The Van Kampen modes being orthogonal to the discrete mode are invisible.

The situation changes in the presence of external nonlinearities, in the particular case of a hollow electron lens. The discrete mode gets submerged into the continuum or "Landau damped"; the stronger the lens the wider the spectrum thus stronger the damping.

Let us point out an important feature of the spectra in Fig. 2: the spectral density maximum – which can be considered as coherent tune – runs away from its original position, $\nu=0$, with stronger hollow e-lens. This means that the coherent tune is determined – contrary to naive expectations – by participation of particles with sufficiently large betatron amplitudes since the hollow lens does not act on particles with amplitudes less than $1\text{-}2\sigma_\perp$.

This run away effect has an important practical consequence that a hollow e-lens is less effective from the point of view of restoration of Landau damping than a bell-shaped lens. Therefore all subsequent analysis will be made for a lens with Gaussian profile.

It would be interesting to look at the spectral density of the head-tail modes with $l\neq 0$. Since the profile of such modes is not known in advance we may take a basis function for it. Figure 3 shows the histogram of $c_{m;4}^2$ for $\nu_s = 0.2\cdot\xi$ in absence of external nonlinearities. Multiple peaks testify that $l_0=4$ basis function is present in eigenfunctions corresponding to $l=2,4,6$ and higher head-tail modes. What is important there is no discrete peaks with $l\neq 0$ – the phenomena called in [15] "intrinsic Landau damping". The damping rate deduced from the peak width is in good agreement with results of multiparticle tracking [16].

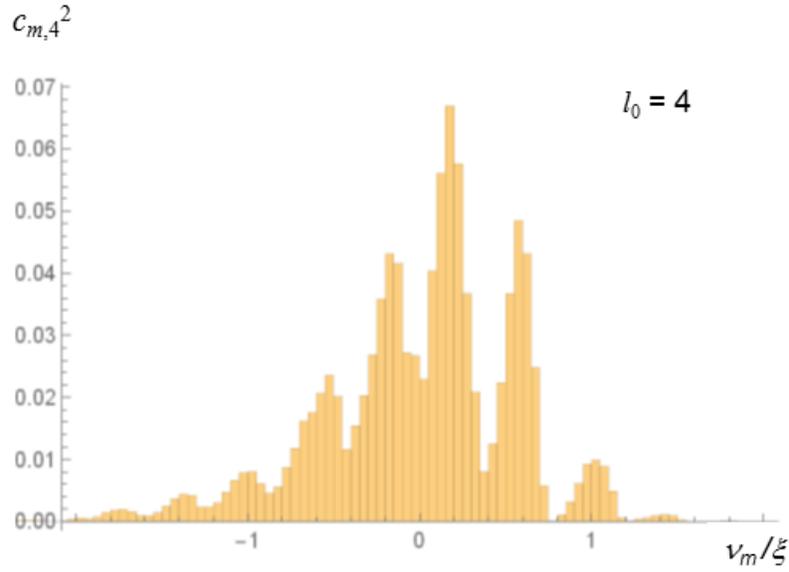

Figure 3. Spectral density of head-tail modes projection on $l_0=4$ basis function.

Since $l\neq 0$ modes are intrinsically damped our main concern is Landau damping of $l=0$ mode. To have a better idea of the damping strength we may employ the stability diagram technique. The required dispersion relation is obtained in following steps.

Let us consider a simple model of external impedance (anti-damper) producing equal kick for all particles in the bunch which is proportional to the whole bunch center-of-mass displacement. Since we are interested in multi-turn phenomena we may spread this impedance evenly around the ring and have for the normalized kick $\Delta$ produced over distance $d\theta$

$$\Delta \equiv \delta p_x \sqrt{\frac{\beta_x}{\varepsilon_\perp}} = -\zeta X d\theta + c.c. \qquad (10)$$

where $\zeta$ is some (complex) constant and $X$ is normalized complex Courant-Snyder variable of the center-of-mass oscillations

$$X = \left\langle \frac{x + i(\beta_x p_x + \alpha_x x)}{\sqrt{\beta_x \varepsilon_\perp}} \right\rangle \tag{11}$$

The second addend in the r.h.s. of eq. (10) vanishes upon averaging over θ so we will ignore it altogether.

The next step is to find the c.o.m. oscillations excited by an elementary kick $\Delta$ delivered at some location, $\theta=\theta_0$. After some manipulations (see e.g. [18]) we get

$$dX = i\Delta(\theta_0) e^{-i\phi_x(\theta) + i\phi_x(\theta_0)} \sum_m e^{-i\nu_m(\theta-\theta_0)} c_{m;0}^2 \tag{12}$$

where the sum is actually the Stieltjes integral (sum over point spectrum and integral over continuum) and $\phi_x = \varphi_x - \nu_x \theta$ is periodic phase function with $\nu_x$ being full bare lattice tune including integer part.

Combining eqs. (10) and (12) we obtain equation

$$X(\theta) = -i\zeta e^{-i\phi_x(\theta)} \int_{-\infty}^{\theta} e^{i\phi_x(\theta')} X(\theta') \sum_m e^{-i\nu_m(\theta-\theta')} c_{m;0}^2 d\theta' \tag{13}$$

which has a non-trivial solution of the form $X(\theta) \sim \exp[-i\phi_x(\theta) - i\nu\theta]$ if $\nu$ satisfies dispersion relation [18]

$$1 = \zeta \sum_m \frac{c_{m;0}^2}{\nu - \nu_m} \tag{14}$$

In absence of external nonlinearities only the discrete mode, $\nu_0=0$, has non-zero spectral coefficient $c_{0;0}=1$ hence eq. (14) solution is $\nu=\zeta$ showing that parameter $\zeta$ is actually the coherent tuneshift. We will use eq. (14) in the subsequent analysis.

Figure 4 presents various shapes of the stability diagram depending on the strength of external nonlinearity (Gaussian e-lens). Please note the transition case of $\Delta Q_e = 3 \cdot \xi$ when there is two boundary points at Re $\zeta= 0$. Of course only the lower one has physical meaning but the upper one indicates where the solution will be for larger $\Delta Q_e$.

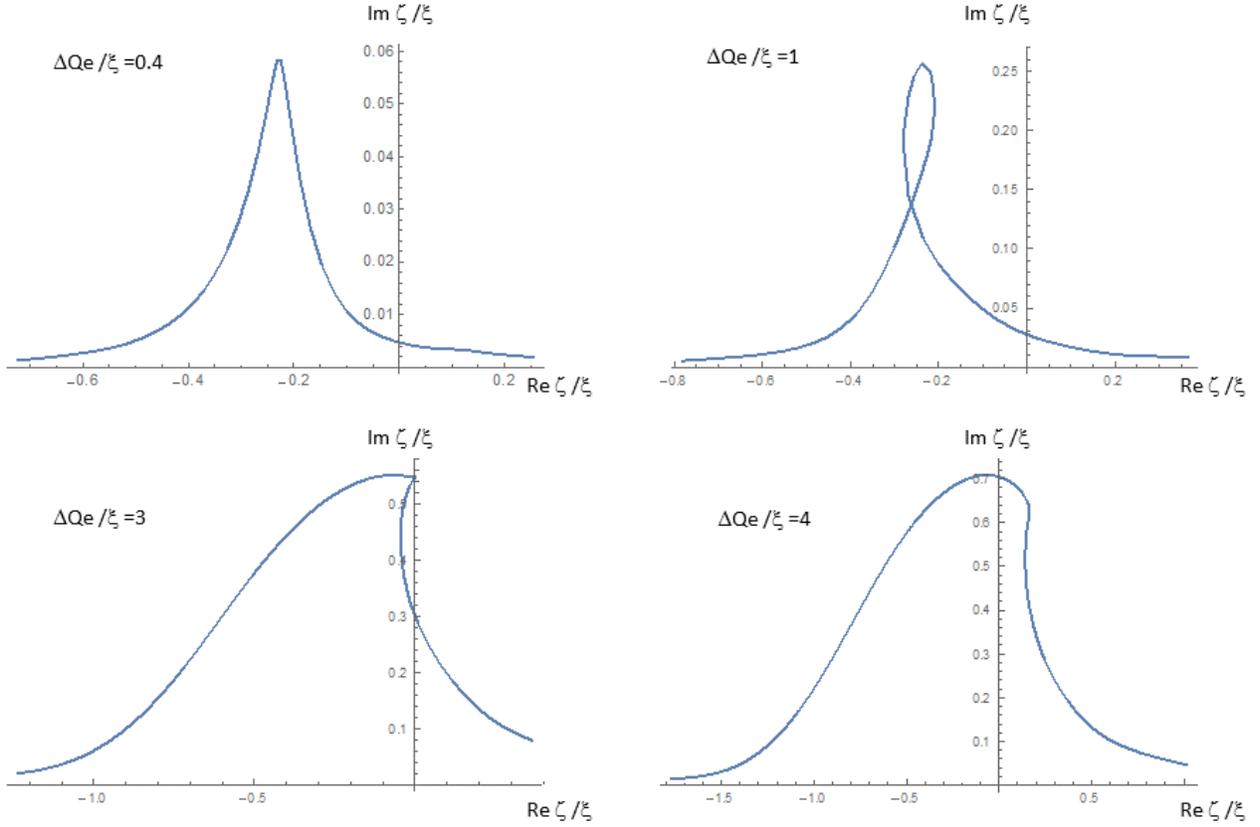

Figure 4. Stability diagrams for different strength $\Delta Q_e$ of Gaussian electron lens and $\nu_s = 0.2 \cdot \xi$.

## 4. ESTIMATIONS OF LANDAU DAMPING FOR ZERO MODE

As a figure of damping efficiency we choose the imaginary part of stability diagram, $\Lambda = \mathrm{Im}\,\zeta$, at $\mathrm{Re}\,\zeta = 0$. This value is addressed below simply as the (Landau) damping rate. In this section we are suggesting qualitative considerations, leading to some estimations of this value at various ranges of parameters, comparing the suggested formulas with the computations of the previous section.

Let us first recollect that for a strong lens, when the maximal values of the tune shifts are such that $\delta\nu_{max} \gg Q_{sc}$, the result was already obtained in Ref. [2]; in this case, $\Lambda = 0.14 \delta\nu_{max}$ for the round Gaussian lens and beam, with their rms sizes identical; note that the rate is independent of the synchrotron frequency. For a flat bunch, as it has been noted above, this formula is true as well, after a slight modification of the numerical coefficient: $0.14 \rightarrow 0.16$.

The opposite case, $\delta\nu_{\max} \ll Q_{sc}$, is especially interesting for the so-called strong space charge regime (SSC, [15]), which assumes that the synchrotron frequency is also small, $\nu_s \ll Q_{sc}$. If so, the beam stability is determined by a competition between two factors, each of them is small compared with the SC: Landau damping and the impedance.

Note first that in this specific case only a small fraction of particles at the bunch tails participate in Landau damping: to make this participation possible, the bunch density should be sufficiently diluted, providing for the lens tune shift to exceed the space charge one. For the Gaussian bunch, it means that the longitudinal offset $\tau$ measured in the r.m.s. units has to be for these Landau particles large enough, $\tau \geq \tau_w$, where the border offset $\tau_w$ is such that $\delta\nu_{\max} = Q_{sc}\exp(-\tau_w^2/2)$, or

$$\tau_w = \sqrt{2\ln(Q_{sc}/\delta\nu_{\max})}, \tag{15}$$

where the subscript $w$ stands for weak lens. The relative number of these particles is then computed as

$$\delta N_w / N = \sqrt{\frac{2}{\pi}} \int_{\tau_w}^{\infty} \exp(-\tau^2/2)d\tau = \text{erfc}(\tau_w/\sqrt{2}) \simeq \frac{\delta\nu_{\max}}{Q_{sc}\tau_w}. \tag{16}$$

The damping rate $\Lambda$ is determined by a product of two factors: by the portion of the Landau particles $\delta N_w/N$ and by their tune spread $\delta\nu$; thus, the damping rate has to scale as $\Lambda \propto \delta\nu_{\max} \cdot \delta N_w/N$. The numerical coefficient can be estimated by taking into account that for $\delta N_w/N = 1$, the damping rate $\Lambda = 0.14\delta\nu_{\max}$. From all that, an estimation for the round Gaussian bunch and the weak lens follows:

$$\Lambda_w = 0.14 \frac{\delta\nu_{\max}^2}{Q_{sc}\sqrt{2\ln(Q_{sc}/\delta\nu_{\max})}}. \tag{17}$$

This expression has a tendency to overestimate the damping rate with growing lens parameter $\delta\nu_{\max}/Q_{sc}$. To prevent that, the expression can be modified as

$$\Lambda_w = 0.14 \frac{\delta\nu_{\max}}{\left(\dfrac{Q_{sc}}{\delta\nu_{\max}} + a_w\right)\sqrt{2\ln(Q_{sc}/\delta\nu_{\max})}}, \tag{18}$$

where $a_w$ is a free parameter to be set by the best fitting with the data. Also, we will slightly modify this formula to prevent zero denominators and negative values under the square root; thus, we may estimate

$$\Lambda_w = 0.14 \frac{\delta v_{max}}{\left(\dfrac{Q_{sc}}{\delta v_{max}} + a_w\right)\sqrt{1 + 2\ln(1 + Q_{sc}/\delta v_{max})}}. \tag{19}$$

To avoid misunderstanding, we would like to stress again, that all the considerations above by no means pretend to anything like strict derivation; instead, they reflect our semi-qualitative reasoning about the structure of the damping rate, where all the numerical coefficients are guessed or fitted. The same logic for the flat bunch would lead to the same formula with the mentioned substitution $0.14 \to 0.16$ for the coefficient in front of the formula. Thus, we may summarize our estimations for the weak-lens damping rate:

$$\Lambda_w = C\delta v_{max} \frac{1}{\left(\dfrac{Q_{sc}}{\delta v_{max}} + a_w\right)\sqrt{1 + 2\ln(1 + Q_{sc}/\delta v_{max})}}; \quad \delta v_{max} < Q_{sc}$$

$$C = \begin{cases} 0.14, & \text{for round bunches} \\ 0.16, & \text{for flat bunches} \end{cases}. \tag{20}$$

For medium and strong lenses, $\delta v_{max} \geq Q_{sc}$, we may guess again that only those particles contribute to the Landau damping whose lens-driven tune shift sufficiently exceeds the space charge tune shift. This suggests the following expression for the damping rate at this regime,

$$\Lambda_s = 0.14 \Delta Q_e \sqrt{\frac{2}{\pi}} \int_{\tau_s}^{\infty} \exp(-\tau^2/2) d\tau = 0.14 \Delta Q_e \, \text{erfc}\left(\tau_s/\sqrt{2}\right) \tag{21}$$

The parameter $t_s$ can be defined by the condition of sufficient excess of the lens over the space charge,

$$\delta v_{max} = a_s Q_{sc} \exp(-\tau_s^2/2), \tag{22}$$

where $a_s$ is the strong lens fitting parameter; presumably $a_s \simeq 2 - 4$, and the subscript $s$ stands for the strong lens. With that,

$$\tau_s = \sqrt{\max\{0, 2\ln(a_s Q_{sc}/\delta v_{max})\}} \qquad (23)$$

Finally, we can try to combine weak and strong lens damping rates in a simplistic way, suggesting a general formula for any lens strength,

$$\Lambda = \Lambda_w + \Lambda_s(1-a_w^{-1}) = 0.14\delta v_{max}\left[\frac{1}{\left(\frac{Q_{sc}}{\delta v_{max}}+a_w\right)\sqrt{1+2\ln(1+Q_{sc}/\delta v_{max})}} + \mathrm{erfc}\left(\tau_s/\sqrt{2}\right)\frac{a_w-1}{a_w}\right] \qquad (24)$$

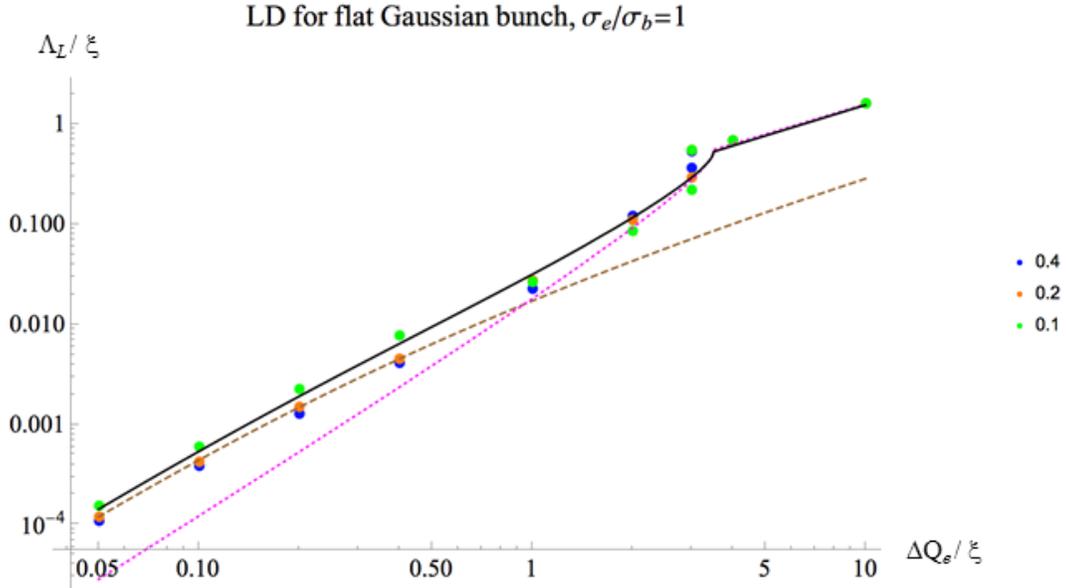

Figure 5: Landau damping rate for a flat Gaussian bunch and $v_s/Q_{sc} = 0.1, 0.2, 0.4$ (green, orange and blue dots). The dots represent direct computations with the Vlasov model of the previous section. The brown dashed line shows the weak lens formula Eq.(20), the magenta dotted line represents the medium-strong lens case of Eq.(21), and the solid black line corresponds to the arbitrary lens case, Eq.(24).

Figure 5 presents results of the damping rate computations with the Vlasov formalism of the previous section compared with the estimation above for the flat bunch with the fitting parameters $a_w = 5.0$ and $a_s = 3.5$. Note that a good agreement is reached within big range of parameters with minimal fitting.

## 5. DISCUSSION AND CONCLUSIONS

Above we have shown feasibility of effective Landau damping with electron lenses in space charge dominated beams. Thinking along practical lines, we suggest to consider the electron lens system for Fermilab's 8 GeV Recycler proton storage ring. Table I lists the parameters of the Recycler and of the electron lens to generate a tune spread $\delta\nu_{max} \approx 0.2$.

Table I: Electron beam requirements to generate the tune shift $\delta\nu_{max} = 0.01$
in the 8 GeV proton beams in the Fermilab's Recycler Ring.

| Parameter | Symbol | Value | Unit |
|---|---|---|---|
| Length | $L_{EL}$ | 2.0 | m |
| Number of elecron lenses | $N_{EL}$ | 2 | one/beam |
| Beta-functions at the e-lens | $\beta_{x,y}$ | 150 | m |
| Current (DC, max) | $I_e$ | 2 | A |
| Electron energy | $U_e$ | 10 | kV |
| e-beam radius in main solenoid | $\sigma_e$ | 0.7 | mm |
| Magnetic field in main solenoid | $B_m$ | 6.5 | T |
| Magnetic field in gun solenoid | $B_g$ | 0.2 | T |
| Tune spread by e-lens (max) | $\delta\nu_{max}$ | 0.02 | |

The technology of the electron lenses is well established and well up to the requirements of the Landau damping in particle accelerators discussed above. Two electron lenses were built and installed in the Tevatron ring [12] at Fermilab, and two similar ones in the BNL's RHIC [13]. They employed some 10 kV Ampere class electron beams of millimeter to submillimeter sizes with a variety of the transverse current distributions $j_e(r)$ generated at the thermionic electron gun, including the Gaussian ones. The electron beams in the lenses are very stable transversely being usually immersed in a strong magnetic field - about $B_g$=1-3 kG at the electron gun cathode and some $B_m$=10-65 kG inside about 2 m long main superconducting solenoids. The electron beam transverse alignment on the high energy beam is done by trajectory correctors to better than a small fraction of the rms beam size $\sigma_e$. The electron lens magnetic system adiabatically compresses the electron-beam cross-section area in the interaction region by the factor of $B_m/B_g \approx 10$ (variable from 2 to 60), proportionally increasing the current density $j_e$ of the electron beam in the interaction region compared to its value on the gun cathode, usually of about 2-10 A/cm$^2$. In-depth experimental studies of the Landau damping with electron lenses are being planned at the Fermilab's IOTA ring [14] which will provide a test bed for exploration of the physics of the space-charge dominated proton beams.

## ACKNOWLEDGEMENTS

*Fermilab is operated by Fermi Research Alliance, LLC under Contract No. DE-AC02-07CH11359 with the United States Department of Energy.*## REFERENCES

including the Gaussian ones. The electron beams in the lenses are very stable transversely being usually immersed in a strong magnetic field - about $B_g$=1-3 kG at the electron gun cathode and some $B_m$=10-65 kG inside about 2 m long main superconducting solenoids. The electron beam transverse alignment on the high energy beam is done by trajectory correctors to better than a small fraction of the rms beam size $\sigma_e$. The electron lens magnetic system adiabatically compresses the electron-beam cross-section area in the interaction region by the factor of $B_m/B_g \approx 10$ (variable from 2 to 60), proportionally increasing the current density $j_e$ of the electron beam in the interaction region compared to its value on the gun cathode, usually of about 2-10 A/cm$^2$. In-depth experimental studies of the Landau damping with electron lenses are being planned at the Fermilab's IOTA ring [14] which will provide a test bed for exploration of the physics of the space-charge dominated proton beams.

## ACKNOWLEDGEMENTS

*Fermilab is operated by Fermi Research Alliance, LLC under Contract No. DE-AC02-07CH11359 with the United States Department of Energy.*

## REFERENCES

1. Shiltsev, in *Proceedings CARE-HHH-APD LHC-LUMI-06 Workshop* (Valencia, Spain, 2006); CERN Yellow Report No. CERN-2007-002 (2007), pp.92-96.
2. V. Shiltsev, Y. Alexahin, A.Burov, and A. Valishev, Landau Damping of Beam Instabilities by Electron Lenses, 2017 (submitted to PRL)
3. A. W. Chao, *Physics of Collective Beam Instabilities in High Energy Accelerators* (John Wiley & Sons, New York,1993)
4. K.Y. Ng, *Physics of Intensity Dependent Beam Instabilities* (World Scientific, 2006)
5. L. Landau, J. Phys. USSR **10**, 25 (1946)
6. J. S. Berg and F. Ruggiero, CERN Report No. CERN-SLAP-96-071-AP, 1996.
7. A. Burov, "Nested Head-Tail Vlasov Solver", Phys. Rev. Accel. Beams **17**, 021007 (2014)
8. V.Shiltsev, *Electron Lenses for Supercolliders* (Springer, 2016)
9. G.Lopez, "Head-on and Long Range Beam-Beam Tune Shift Spreads in the SSC", SSCL-Preprint-304 (1993); Proc. IEEE PAC'93 (Washington, DC, 1993), pp.3467-3469